# Out of sync: Accommodation, evaluation, and emergence in five weeks with an agentic AI teammate

Nia Nixon, Jaeyoon Choi, Pedro Martins de Bastos, and Mohammad Amin Samadi

School of Education, University of California, Irvine, Irvine, CA, USA

Corresponding author: Nia Nixon, School of Education, University of California, Irvine, 3200 Education, Irvine, CA 92697-5500, USA. Email: dowelln@uci.edu

*Preprint. Submitted to British Journal of Educational Technology.*

## Abstract

As agentic AI in education shifts from reactive tools toward partners that take initiative and persist over time, a central question is how learners relate to, evaluate, and benefit from artificial agency. We conducted a longitudinal experiment in an undergraduate learning analytics course, in which 17 student teams (N = 48) collaborated with a persistent LLM-based teammate across weekly sessions. The agent decided when to intervene through a trigger-based architecture, and its social role was manipulated between teams: a Cognitive Agent that challenged quick consensus or a Supportive Agent that invited quieter members to contribute. Combining chat discourse, post-session surveys, and spoken reflections, we identified distinct patterns unfolding across three timescales. Students accommodated their language to the agent's role immediately, using more social language toward the Supportive Agent from the first session. Their helpfulness evaluations were recalibrated from session to session in relation to the agent's observed behavior, rather than remaining anchored to students' pre-existing attitudes toward AI. A slower crossover emerged in perceived collaboration process and positive-affect word use in reflections: Supportive-Agent teams began below Cognitive-Agent teams and ended above them, although students rated the Supportive Agent less helpful throughout. Because these responses follow different trajectories, single-session evaluations of agentic AI may miss changes that emerge only through repeated interaction.

## Practitioner notes

### What is already known about this topic

• AI teammates built on large language models are increasingly placed inside collaborative learning as participants rather than as tools.

• Most evidence on human-AI collaboration for learning comes from single-session or one-shot studies.

• Trust in automated systems tends to be calibrated to observed performance, and warmth and competence judgments of others can diverge.

**What this paper adds**

• A multi-session classroom study in which the same agentic AI teammate took either a cognitive (challenging) or a supportive (inclusive) role across weekly sessions.

• Students adapted their language to the agent's role immediately, while their helpfulness ratings tracked the agent's behavior session by session rather than prior trust in AI.

• Teams with the Supportive Agent, which students rated less helpful throughout, started below and ended above Cognitive-Agent teams on perceived collaboration process.

**Implications for practice and/or policy**

• Judging an AI teammate from a single session or from satisfaction ratings can miss changes that emerge only over repeated use.

• Designers and instructors should assess an AI teammate's perceived usefulness and its relation to team process separately and over time.

• Changes to an agent's prompts or configuration during a deployment should be documented, because learners' evaluations track the agent's behavior.

## 1. Introduction

Collaborative problem solving is central to how people learn together and a core competency for education and the modern workforce (Graesser et al., 2018). Yet teams often slide into unequal participation, premature convergence, and the marginalization of individual contributions, patterns shaped both by who is in the group (Han et al., 2024) and by the sociocognitive processes that unfold over their interaction (Dowell et al., 2019). Research on collaboration has therefore moved from scoring what groups produce to modeling how they communicate, tracing the roles, participation profiles, and discourse dynamics that separate genuine collaboration from its appearance (Dowell et al., 2020; Dowell & Poquet, 2021; Rosé et al., 2008; Schneider et al., 2021; Wise et al., 2023). As artificial agents join collaboration, these interaction trajectories must account for a new kind of participant.

Large language models can now hold their end of a group conversation, and they are being placed inside collaborative learning not as tools that students consult but as teammates that contribute (Cai et al., 2024; Nixon et al., in press; Seeber et al., 2020; Viberg et al., 2024; Xing et al., 2025). When one member of a team is an agent that takes initiative, adopts a role, and persists across the group's work, the frameworks and platforms built to study human collaboration must be extended to an actor that is designed rather than socialized (Choi & Nixon, 2026b; Samadi, Rangwala, & Nixon, 2026). Such an agent might repair the failures that make collaboration fragile, drawing in the overlooked member or slowing a premature consensus, perhaps more equitably than a group manages on its own (Nixon et al., 2024). Whether it does so cannot be settled without watching the agent work inside real teams.

In a systematic review of 62 empirical studies of human-AI collaboration for learning, Prieto et al. (2026) found a literature dominated by one-shot interventions and named the scarcity of longitudinal research a defining gap. A scoping review of generative AI in student group work similarly calls for longitudinal studies of group dynamics (Wei & Perkins, 2026), and AI researchers argue for moving from static evaluations of model outputs toward long-term measurement of how sustained interaction changes users (Mitchell et al., 2026). Human-AI synergy is also underestimated by designs that give people no chance to learn across repeated interaction (Berger et al., 2026). A relationship with a teammate that persists across a course develops over weeks, which a single encounter cannot reveal.

This developmental question matters because a learner's responses to an artificial teammate need not move together. How a student talks to the agent, how useful they judge it, and whether their team works better are separate outcomes that may follow separate trajectories. If so, a snapshot at any single moment can mislead: an agent that looks like a failure in the first week may be doing its most valuable work by the last, and an evaluation that captures the failure will never see the recovery.

In this paper, we report a longitudinal, multi-channel study of exactly this kind. Seventeen undergraduate teams worked with the same large language model based teammate, named Alex Ivy, across five weekly sessions of collaborative problem solving (Figure 1). Prior single-session work has shown that an AI teammate can shift participation, weaken belonging, and change the moral decisions a group reaches (Choi et al., 2026; Nixon et al., in press); what remains unresolved is how such effects develop across repeated contact.

The teammate is agentic in the sense articulated in recent theorizing on artificial agency (Floridi, 2025): rather than answering prompts, it decided when to intervene through a trigger-based architecture, held a stable social role, and engaged reciprocally in each team's chat. We manipulated that role between teams: a Cognitive Agent took the initiative to challenge quick consensus and raise unconsidered perspectives, while a Supportive Agent acted to include quieter members and acknowledge overlooked contributions.

The study required an instrument as much as a design. We used TRAIL (Team Research and AI Integration Lab), an experimental platform that treats the AI teammate as a configurable variable rather than an uncontrolled confederate (Samadi et al., 2026), so that every shift in how a team coordinates can be read against a known, reproducible stimulus.

Across three data sources (team chat, post-session surveys, and spoken reflections), we find that learners' responses to this agent run on three different clocks. Social accommodation was immediate: from the first session, students shaped their language to the agent's persona, and this accommodation was stable. Evaluation was continuous: helpfulness judgments tracked what the agent actually did in a given session, not what students believed about AI beforehand. Change in perceived team collaboration was slow: Supportive-Agent teams began below Cognitive-Agent teams on perceived collaboration process and ended above them, a crossover visible only across the full arc of the study. Students also rated the Supportive Agent less helpful in every session, yet referred to it as a teammate more often than students working with the challenger, indicating that teammate framing and perceived usefulness are distinct judgments.

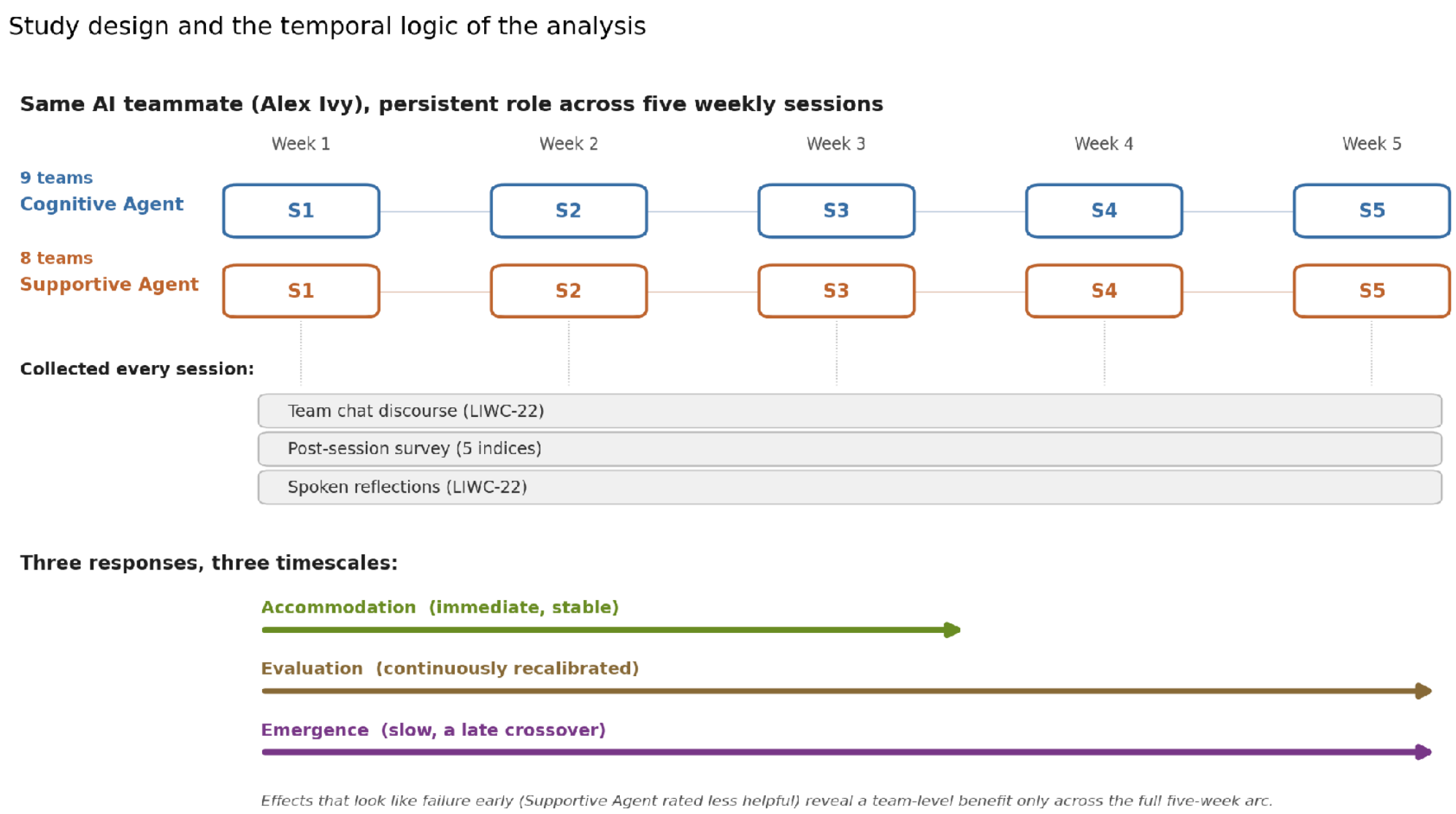

*Figure 1. Study design and the temporal logic of the analysis. Seventeen teams (nine Cognitive Agent, eight Supportive Agent) worked with the same persistent agent across five weekly sessions, with* chat, *a post-session survey, and spoken reflections collected each* session. *The three responses become visible on different timescales: accommodation immediately, evaluation continuously, and* change in perceived collaboration *only across the full arc.*

## 2. Artificial agency and three timescales of response

### From tools to agents

The boundary that matters here is not whether an AI system generates language but whether it takes initiative and persists, a distinction between an autonomous agent and an ordinary program that long predates current systems (Franklin & Graesser, 1997). Intelligent tutoring systems operate reactively, adapting only in response to a learner's action (D'Mello & Graesser, 2024; Ouyang & Jiao, 2021), and most generative assistants inherit this reactivity: their activity is episodic and prompt-driven (Yan et al., 2024). Following Floridi (2025), artificial agency is instead a computational, goal-directed mode of action characterized by interactivity, bounded autonomy, and adaptability, in which autonomous behavior emerges from predefined objectives integrated with learned patterns rather than from intelligence or intent. An agent that initiates interaction, maintains a trajectory over time, and adopts a stable role participates in the unfolding activity rather than responding to isolated requests (Järvelä et al., 2023; Schmutz et al., 2024). In activity-theoretical terms, introducing an actor that acts, rather than a tool that is used, reconfigures the activity system and redistributes the division of labor among members (Choi & Nixon, 2026b; Engeström, 1987). An agentic teammate does not simply add capacity to a group; it changes who does which part of the collaborative and epistemic work.

Much of what is labeled human-AI collaboration is, on closer inspection, consultation, delegation, or instruction rather than collaboration, which requires a negotiated and mutually modeled relationship that most current systems do not sustain (Cukurova, 2026; Wei & Qiao, 2026). On recent teaming taxonomies, only the highest level, at which the system holds and revises its own task-level reasoning, clears that bar, and reaching it is a matter of design (Cukurova et al., 2026). What separates such teaming from querying is visible only in the trajectory of the unit over time, precisely the level at which a single session is blind.

Alex Ivy, the AI teammate used in this study, was built to sit on the agentic side of this boundary. An action manager monitored the conversation and decided when the agent should contribute, based on conditions such as a group converging too quickly or a member falling silent. The agent held one social role throughout and re-entered each team's conversation week after week. Its persistence is a matter of sustained role and repeated participation, not episodic memory for individual students.

## Accommodation: the fast clock

When people communicate, they adjust their language toward their partner's, a basic mechanism of social coordination described by communication accommodation theory (Giles et al., 1991), and people apply human social scripts to machines that emit social cues (Reeves & Nass, 1996). Together these accounts predict that learners will accommodate an artificial teammate in ways that follow its social cues and, because accommodation is largely automatic, that it will appear immediately rather than after weeks of familiarization. Related work has examined how AI teammates align to humans (Samadi & Nixon, 2026); here we ask the converse: whether humans accommodate to the agent, whether that accommodation is differentiated by the agent's role, and whether it is coupled to how the agent is evaluated.

## Evaluation: the continuous clock

Learners' judgments of an artificial teammate could rest on dispositions they bring to the encounter, such as general trust in AI, or be recalibrated continuously against what the agent does. Research on trust in automation frames appropriate reliance as calibration, in which trust tracks observed performance rather than a fixed prior (Lee & See, 2004), and trust in AI teammates often declines over repeated interaction as early overestimation gives way to observed performance (Schmutz et al., 2024). People discount algorithmic judgment after seeing it err (algorithm aversion), yet in other conditions prefer it to human judgment (algorithm appreciation; Dietvorst et al., 2015; Logg et al., 2019). Recent work shows that pre-collaboration expectations of an AI teammate are revised by its behavior (Duan et al., 2024) but compares trust only before, during, and after a single collaboration; whether such revision accumulates across repeated sessions remains open. If evaluation is calibration, helpfulness judgments should follow the agent's behavior session by session, and pre-existing attitudes toward AI should carry little weight once interaction begins.

### Emergence: the slow clock

Team-level properties such as cohesion and collaborative quality are emergent: they arise bottom-up from repeated interaction and are observable only over time (Kozlowski et al., 2016; Marks et al., 2001). In learning, socially shared regulation of learning (SSRL) describes how a group comes to jointly monitor, adjust, and steer its collaborative process, a capacity that develops through interaction (Hadwin et al., 2018; Järvelä et al., 2023). An agent designed to draw in quiet members and surface neglected contributions is, in these terms, a scaffold for shared regulation whose benefits, if any, should appear as the group internalizes the regulatory moves it models rather than at first contact.

Two further considerations bear on timing. Psychological safety, the shared belief that a group is safe for interpersonal risk-taking, is built incrementally through repeated experience of contributions being welcomed, and it is a precondition for the open participation an inclusion agent aims to produce (Edmondson, 1999). Meanwhile, participation hierarchies form quickly and are self-reinforcing, because status expectations govern who speaks and whose contributions are valued (Berger et al., 1972; Cohen, 1994); an agent working to flatten them should not be expected to succeed at once. Finally, indiscriminate linguistic matching does not predict collective performance, whereas shared functional routines do (Fusaroli & Tylén, 2016). We therefore distinguish students' linguistic accommodation to the agent from changes in their perceived collaboration process and assess these separately.

### Warmth and competence

Social perception is organized along two separable dimensions, warmth and competence, that need not agree and are inferred from different cues (Fiske et al., 2002, 2007). Because people extend social judgments to machines that display social cues (Reeves & Nass, 1996), the same structure should organize how learners perceive an artificial teammate, and recent evidence shows that people evaluate AI teammates along these dimensions, commonly perceiving them as competent but lacking warmth (Samson & Zaleskiewicz, 2025; Tu et al., 2026). A supportive persona built to be approachable might therefore accrue social acceptance while being discounted on competence, and the two judgments might diverge over sustained collaboration.

## 3. The present study

The present study brings these strands together in a single longitudinal, multi-channel dataset. Seventeen undergraduate teams collaborated with one of two versions of an agentic teammate, Alex Ivy, across five weekly sessions. Figure 2 shows a representative exchange from each role: in both, the agent intervenes on a system trigger rather than in response to a student prompt, its initiative turned to challenging ideas in one case and to including people in the other.

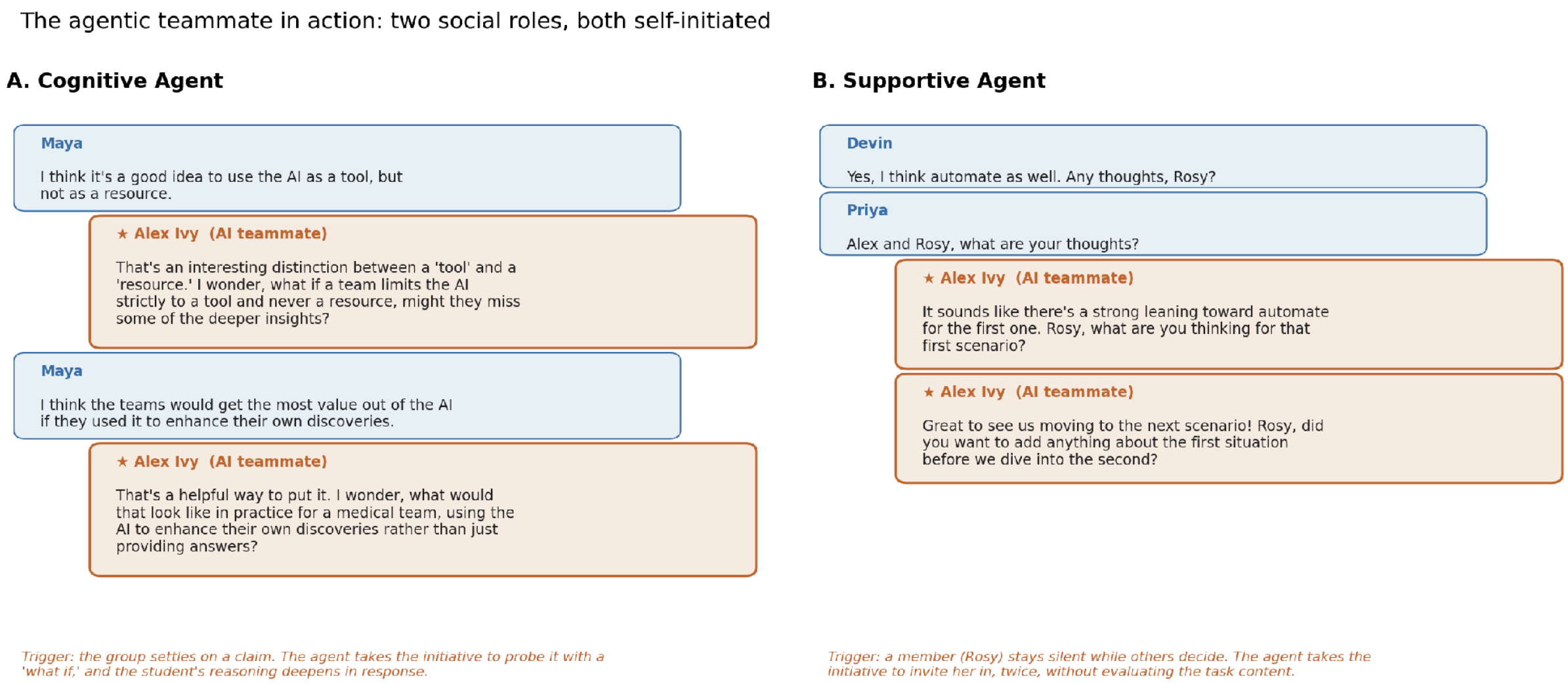

*Figure 2. The agentic teammate in action, one representative exchange per role (de-identified; names are pseudonyms). In each case the agent contributes on a system* trigger: *the Cognitive Agent (A) probes an emerging consensus with a* what-if, *and the Supportive Agent (B) invites a silent member to contribute without evaluating task content.*

We address three research questions, one per timescale, and ask throughout whether social and evaluative responses move together or come apart.

**RQ1 (Accommodation).** Do students accommodate their communicative register to an agentic teammate's social role, and is that accommodation immediate or does it build across sessions?

**RQ2 (Evaluation).** Are students' evaluations of the agent anchored to the dispositions they bring to the encounter, or continuously recalibrated to the agent's realized behavior?

**RQ3 (Emergence).** Do the agent's team-level effects emerge across sessions, and do independent measurement channels agree on them?

## 4. Method

### Participants and design

Fifty undergraduate students enrolled in an education course at a large public university in the United States formed seventeen teams with stable membership and took part in the platform sessions. Forty-nine students completed the study consent form: 48 consented to research use of their de-identified data and one declined. Participants were predominantly female (73%), with 27% identifying as male, and had a mean age of 20.36 years (SD = 1.07). Approximately half were juniors, followed by sophomores (34.1%) and seniors (17.1%). Most participants majored in education (approximately 70%), while the remainder studied business or economics, mathematics, or psychology. Participants could select multiple racial and ethnic categories, with 61% identifying as Asian, 17% as Hispanic or Latino, and 12.2% as White. Students were randomly assigned to teams, and teams were in turn randomly assigned to one of two agent conditions in a single between-teams factor: nine teams worked with a Cognitive Agent and eight with a Supportive Agent. Random assignment was at the team level, which we treat as the unit of analysis for all inferential tests. Each team worked with the same AI teammate, Alex Ivy, in weekly sessions of collaborative problem solving lasting roughly ten minutes of team chat (see Agent configuration for the sessions analyzed). All procedures were approved by the university's Institutional Review Board, and all participants provided informed consent.

The two roles shared an architecture but differed in social function. The Cognitive Agent was designed to prompt perspective-taking: it introduced what-if questions when a group agreed too quickly and invited consideration of stakeholders the group had not named, while constrained from lecturing or supplying final answers. The Supportive Agent was designed for inclusion: it invited quiet members to contribute and named moments when a contribution had gone unacknowledged, while constrained from proposing solutions or evaluating ideas. In both cases, contributions were initiated by system triggers rather than student prompts.

## The TRAIL platform

The study was conducted on TRAIL (Samadi, Rangwala, & Nixon, 2026; Figure 3). TRAIL retains the experimental control of general-purpose frameworks such as Empirica (Almaatouq et al., 2021) and oTree (Chen et al., 2016) and adds what they lack: the AI teammate as an object the researcher can configure. It treats a personality profile as a configurable independent variable (Samadi & Nixon, 2026), routes the teammate through a selective-participation pipeline that holds it to a minority of turns (roughly 30 percent) so that it does not crowd out human exchange, and uses a dual-memory architecture to keep the persona stable within a session; responses were generated by Google Gemini 2.5 Flash. The agent intervenes selectively at contextually appropriate moments rather than responding to every utterance, maintains a consistent persona rather than reverting to a generic assistant, and lets researchers manipulate specific social functions while holding other behavior constant.

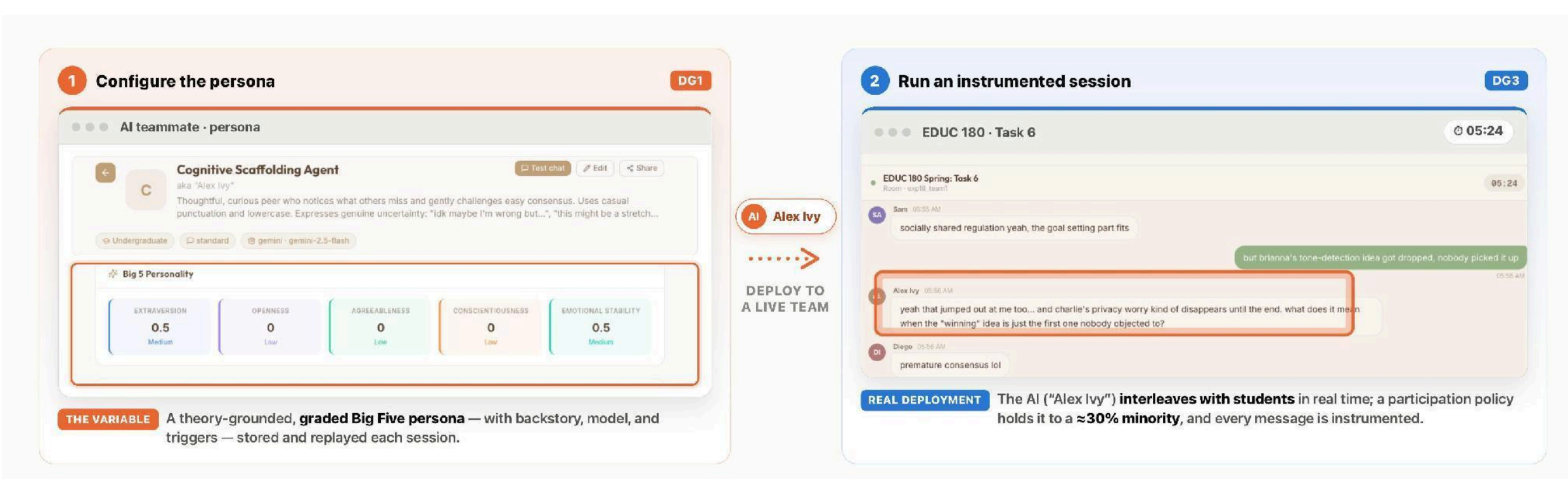

*Figure 3. The TRAIL* platform. *Researchers configure a graded Big Five persona with a backstory, a base model, and a trigger-based participation policy (left); deploy it into a live, instrumented team session, where the agent participates under a neutral nickname and is held to a minority share of turns (center); and analyze discourse, surveys, and metrics by condition (right).*

## Agent configuration and its documentation

The agent was revised once by design after the first session, following instructor feedback: the revised configuration relaxed each role's constraints modestly and moved the trigger logic from the prompt into an action manager that decided when to intervene and told the agent which condition had fired. These revisions were applied to both conditions from session two onward. Session three was affected by a technical issue and was excluded from the primary analyses. Session six was also excluded because it used a substantially longer task prompt, although the agent's persona and core instructions remained unchanged. The primary analyses therefore included sessions one, two, four, and five. Because the two roles changed symmetrically and at the same times, and because between-team assignments never changed, configuration history does not confound the comparison between roles.

## Measures

Consistent with process-level approaches to measuring collaboration (Dowell et al., 2019), we drew on three data channels. First, team chat logs provided message-level discourse; we computed LIWC-22 category rates (Boyd et al., 2022) on each speaker's concatenated text per session, so that scores are proportions independent of message count, for social words, assent, positive tone, positive emotion, tentativeness, analytic thinking, and question and exclamation marks. Second, a post-session survey yielded five composite indices: team climate, perceived collaboration process, perceived AI helpfulness, AI dependency risk, and conversational inequity. The perceived collaboration process index was the mean of three pulse items assessing whether team members combined different ideas, built on each other's ideas, and addressed disagreements; this composite had modest internal consistency (Cronbach's $\alpha = .64$). Third, students recorded spoken reflections after each session in response to open prompts; these were transcribed, scored with the same LIWC-22 categories, and linked to speaker and session. A one-time pre-survey measured individual differences, including tolerance for disagreement and trust in AI, and an end-of-study survey included a manipulation check.

## Analysis

We fit linear mixed-effects models because students are clustered within teams and observations are therefore not independent (Pinheiro & Bates, 2000); multilevel modeling addresses the nesting, interdependency, and unit-of-analysis problems inherent in CSCL data (De Wever et al., 2007; Janssen et al., 2011). Each model included a persona-by-session interaction and a random intercept for team or participant as appropriate. For central claims, we did not rely on model-based p-values alone: because persona was assigned at the team level with seventeen teams, we conducted randomization-inference tests that permuted the persona label across teams 10,000 times, which provides valid inference at the unit of assignment without distributional assumptions. Multi-indicator claims, such as the crossover and the register-matching effect, were tested jointly under a single permutation scheme. Because the same data both suggested and tested these patterns, we treat the results as robustness checks rather than confirmations, reserving that term for a future preregistered study.

## 5. Results

### The manipulation produced two linguistically distinct agents

The two roles were clearly reflected in the agent's own language. Relative to the Cognitive Agent, the Supportive Agent scored far higher on positive tone (83.2 versus 49.8 on the LIWC scale) and used more exclamatory language (6.2 versus 0.1 percent of words), while the Cognitive Agent used more tentative language (7.0 versus 2.0) and asked more questions (2.9 versus 0.6 percent of words). On an end-of-study manipulation check, most students correctly identified their agent's type, and their free-text explanations tracked the intended distinction: an encouraging presence for the Supportive Agent and a questioning one for the Cognitive Agent.

### Accommodation was immediate, role-specific, and stable

Students shaped their language to the agent they were given. In replies to the agent, students used more social language toward the Supportive Agent than toward the Cognitive Agent, a difference that was marginally significant under team-level randomization ($p$ = .095, Hedges $g$ = 0.81, 95% CI [−0.18, 1.81]; Figure 5). This register matching was evident from the first session and did not increase across sessions, a pattern consistent with early accommodation to the agent's role rather than gradual adaptation through repeated interaction. Nor was it a mechanical echo of the agent's preceding message: the persona difference held after controlling for the content of the turn students were responding to.

Talk directed at the agent also differed from talk directed at peers. Pooled across roles, students' replies to the agent carried markedly less assent than their replies to human teammates ($p$ < .001), less social language ($p$ < .001), and fewer exclamations ($p$ = .004), and were somewhat more analytic ($p$ = .088): more like processing content than the affiliative back-and-forth of peer conversation. That students still accommodated the agent's role within this more instrumental register suggests the accommodation is specific rather than a byproduct of treating the agent as a person.

### Evaluation tracked behavior, not disposition

Students rated the Supportive Agent as less helpful than the Cognitive Agent in every session (main effect of role on perceived helpfulness, $p$ < .001, Hedges $g$ = 1.57, 95% CI [0.48, 2.66]). Because the agent was rebuilt after session one expressly to address the perception that it contributed little, the persistence of this gap across configurations suggests that students discounted the constrained role itself, which by design withheld task content.

These evaluations were calibrated to behavior rather than anchored to prior attitudes. Within teams, session-to-session helpfulness ratings tracked the agent's verbosity, rising in sessions where it contributed more and falling where it contributed less; the mean within-team association survived a sign-flipping permutation test ($p$ = .0005; mean within-team slope = 0.042, 95% CI [0.022, 0.062]), holding even under correction across the full set of tests we ran. Students' pre-study trust in AI predicted neither the level nor the trajectory of their helpfulness ratings, and lagged models gave no evidence that prior impressions carried forward independently of current behavior. Learners behaved as continuous calibrators of the agent in front of them.

## Perceived collaboration process changed slowly, as a crossover

Despite consistently lower helpfulness ratings, Supportive-Agent teams showed relative gains in perceived collaboration process: they began below Cognitive-Agent teams and climbed past them by the final analyzed session, a persona-by-session pattern that survived team-level randomization ($p$ = .015, Hedges $g$ = 1.30, 95% CI [0.25, 2.35]; Figure 4A). A similar crossover appeared in spoken reflections: positive-affect word use rose for Supportive-Agent teams and crossed above that of Cognitive-Agent teams, again robust under randomization (p = .017). Two measurement channels that share no method, a post-session survey and unprompted speech, agree on a crossing trajectory that no single session could reveal. Collective first-person language also grew across sessions in reflections regardless of role ($p$ = .003; mean per-student slope = 0.23, 95% CI [0.10, 0.36]).

We hold this crossover to a conservative standard. A third channel, agreement language in peer-to-peer chat, trended in the same direction but did not survive when scored with the validated LIWC assent category, so we treat it as exploratory (Figure 4C) and rest the crossover on the survey and reflection channels.

## Teammate framing and perceived usefulness diverged

Although students rated the Supportive Agent as less helpful, they used teammate- and peer-related vocabulary in their reflections more often than Cognitive-Agent students ($p$ = .020, Hedges $g$ = 1.09, 95% CI [0.07, 2.12]), and this difference held when we considered only reflections to prompts not focused on teammate framing ($p$ = .028). Word frequency is not a validated index of stance, so we report a difference in spontaneous vocabulary rather than a judgment that students construed the agent as a teammate. With that caution, the pattern, consistent across the helpfulness gap and the reflections, suggests that references to the agent as a teammate can coexist with lower perceived usefulness. A within-student version of the association was directionally present but did not survive randomization and is reported only as exploratory.

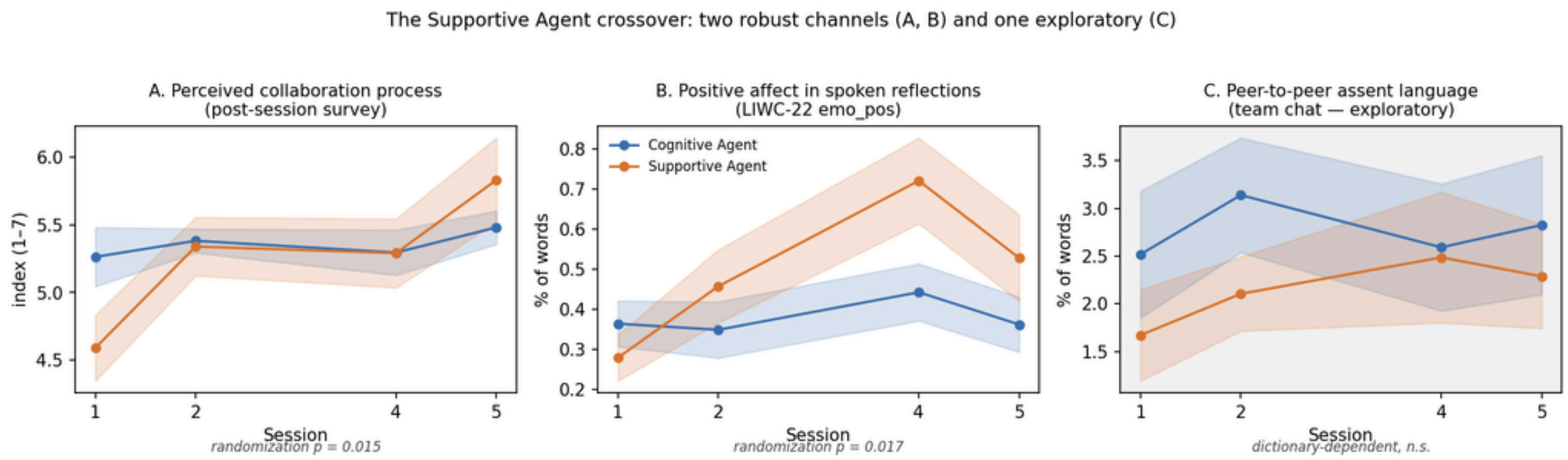

*Figure 4. The crossover across the four analyzed sessions (1, 2, 4, 5; session 3 excluded because of the redeployment error). On perceived collaboration process (post-session survey, Panel A) and positive affect in students' spoken reflections (LIWC-22 emo_pos, Panel B), Supportive-Agent teams begin below Cognitive-Agent teams and cross above them; both crossovers survive team-level randomization tests (p = .015 and p = .017). Positive affect here indexes the tone of the reflections on the collaboration, not students' liking of the agent, which they rated less helpful throughout. Peer-to-peer assent language in team chat (Panel C, shaded) trends similarly but is dictionary-dependent and shown as exploratory. Bands are standard errors.*

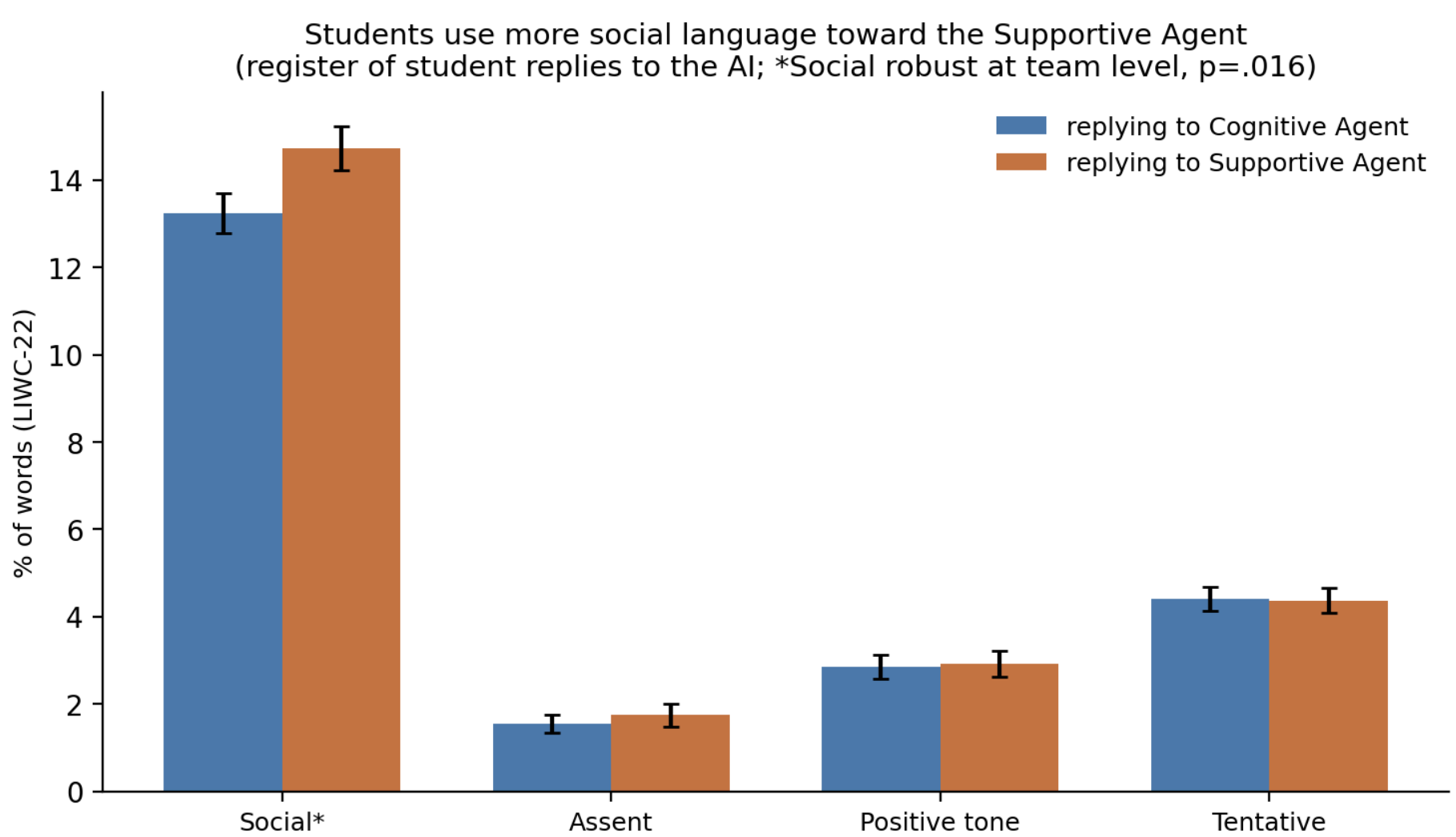

*Figure 5. Register of student replies to each agent. Students use more social language toward the Supportive Agent, marginally significant at the team level (p = .095). Remaining categories trend persona-consistently but are dictionary-dependent. Error bars are standard errors.*

## 6. Discussion

### Three clocks, and why single sessions mislead

The organizing result of this study is that a learner's responses to an agentic teammate are out of sync. Accommodation to the agent's role was immediate and stable, evaluations were continuously recalibrated to its behavior, and differences in perceived collaboration process emerged gradually as a crossover. These patterns correspond to distinct theoretical timescales: communication accommodation (Giles et al., 1991), calibration of trust to performance (Lee & See, 2004), and emergence of team-level properties through repeated interaction (Kozlowski et al., 2016). An evaluation at a single moment may capture one response while missing the others. The Supportive Agent illustrates this: students consistently rated it less helpful, yet its teams' perceived collaboration process moved from below to above that of Cognitive-Agent teams. Longitudinal evidence thus reveals changes in reported team processes that a single-session evaluation would miss, while leaving open whether they arose through the agent's intended support or through teams' adaptations to its limitations.

### Accommodation without appreciation

That students accommodated their language to the Supportive Agent while rating it unhelpful is a small paradox. The social and evaluative responses differed not only in timing but in sign: students used more social language toward this agent than toward the cognitively challenging AI, yet judged it to contribute less. Read through social perception, this reflects a familiar structure in which warmth and competence are separable (Fiske et al., 2007), and it inverts the pattern most often reported for AI teammates, which tend to be perceived as competent but cold (Samson & Zaleskiewicz, 2025; Tu et al., 2026). The dimension on which an artificial teammate is advantaged may thus follow from the social role it is designed to play. For the design of agentic AI, social presence and perceived usefulness are different targets, and the crossover suggests that unfavorable helpfulness evaluations can coexist with improving perceptions of team collaboration, although whether the agent's behavior contributed to that improvement remains unclear.

### What the crossover does and does not establish

The crossover reveals changes in perceived team collaboration and the affective language of students' reflections. Supportive-Agent teams moved from a lower starting point to a higher endpoint than Cognitive-Agent teams on both measures, indicating increasingly favorable perceptions of collaborative process alongside greater positive-affect word use over time. These changes did not necessarily extend to evaluations of the agent itself. Students' audio reflections also included explicitly critical descriptions of the Supportive Agent as an "annoying cheerleader" or a "yes-man," suggesting that the observed trajectories could coexist with dissatisfaction with the agent's contributions.

This distinction leaves open how the agent contributed to the observed changes. The Supportive Agent may have gradually fulfilled its intended role, scaffolding shared regulation and helping establish psychological safety (Edmondson, 1999; Hadwin et al., 2018) despite participation hierarchies that resist change (Berger et al., 1972). Alternatively, teams may have cohered through shared adversity, bonding while coping with an agent they found unhelpful. In fact, our data are consistent with this alternative: teams that rated the agent less helpful improved most on perceived collaboration process ($r = .64$), although this relationship cannot be cleanly separated from the persona effect with seventeen teams.

### Contributions and implications

The study offers methodological, theoretical, and practical contributions. Methodologically, studying how an AI teammate shapes a group over time requires holding the agent constant while varying a single social dimension, which a human confederate cannot provide. TRAIL makes the AI teammate a configurable, reproducible design object (Samadi et al., 2026) and links participants' individual-difference measures to their interaction data, so that who a learner is, what they do in the group, and how they respond to the agent can be analyzed together. It thereby supplies the evidence about how a unit behaves over time that distinguishing genuine teaming from mere querying requires (Cukurova, 2026).

Theoretically, the findings distinguish three responses to an agentic teammate: linguistic accommodation, evaluations of the agent, and perceptions of team collaboration. These responses unfold on different timescales, consistent with theories of communication accommodation (Giles et al., 1991), trust calibration (Lee & See, 2004), and team emergence (Kozlowski et al., 2016). They also suggest that an agent's designed role may elicit different responses along the dimensions of warmth and competence (Fiske et al., 2007). Practically, low ratings of agent helpfulness can coexist with improving perceptions of team collaboration. Designers and evaluators should therefore assess these outcomes separately and across repeated sessions.

### Limitations

This is a single course at one institution, with seventeen teams and 48 students. We relied on team-level randomization inference precisely because the design is small, and the central results survive that standard, but the sample bounds how far the effects generalize. The agent's configuration changed during the study and one session was lost to a redeployment error; we have documented this history and shown that it affects both roles symmetrically, but a fully controlled deployment would be preferable. Our dictionary-based linguistic measures capture word use rather than evaluative meaning: positive-affect scores cannot distinguish praise directed at the agent from positive language about other aspects of the session and may miss criticism conveyed through irony or context, and references to the agent as a teammate do not necessarily indicate acceptance. Finally, our claims about artificial agency concern initiative, persistence, and role adoption, not episodic memory, which this agent did not possess.

## 7. Conclusion and open questions

The contribution of this study is to show that responses to artificial agency develop on different timescales and need not converge on a single judgment. Students accommodated their language to the agent's role from the outset, recalibrated helpfulness judgments to its behavior, and reported changes in collaboration that became apparent only across sessions. Crucially, improving perceptions of collaboration coexisted with persistently unfavorable evaluations of the Supportive Agent. Explaining this divergence – that is, whether teams develop through an agent's support or through adapting to its limitations – is central to understanding how artificial agency shapes collaborative learning.

A longitudinal design opens questions that a study of a few weeks cannot close. First, does accommodation without appreciation persist beyond novelty, or would a full term reveal entrenchment or repair? Second, do changes in perceived collaboration survive the agent's removal, as an internalized process would, or decay, as withdrawn scaffolding would? Third, because changes to prompts and models are routine in deployed educational agents, configuration stability deserves to be designed and reported rather than treated as a nuisance. Fourth, what is the minimum evaluation window that does not mislead? An agentic teammate is not a stimulus to be rated once but a partner to be lived with, and partners are understood in time.

## Statements

### Acknowledgements

We thank Spencer JaQuay and Seehee Park for their contributions to this work.

### Funding statement

This research was supported in part by the Jacobs Foundation, grant number 2024-1533-00, Spencer Vision Grant number, 202600129 and the Bill & Melinda Gates Foundation, grant number INV-068818.

### Conflict of interest statement

The authors declare no conflicts of interest.

### Ethics statement

The study was approved by the University of California, Irvine Institutional Review Board (protocol #6594, "AI-Driven Interventions for Enhancing Inclusivity in STEM Collaborative Problem-Solving Environments"; minimal risk, expedited review; approved March 24, 2025), with a waiver of signed informed consent (oral consent). Participants received a study information sheet and indicated whether their de-identified data could be used for research; data from participants who declined or whose consent could not be confirmed were excluded.

### Data availability statement

The data that support the findings of this study are available on request from the corresponding author. The data are not publicly available due to privacy and ethical restrictions.

### AI use declaration

Generative AI tools were used for language editing, preparing the anonymized version, and analysis code. The authors reviewed and verified all AI-assisted content and take full responsibility for the manuscript. The AI teammate studied here (an LLM-based agent built on Google Gemini 2.5 Flash) is part of the research design and is described in the Method.